\documentclass{webofc}
\usepackage[varg]{txfonts}   % Web of Conferences font
\usepackage{subcaption}

\begin{document}
\title{Study of laser-beam arrival time synchronization towards sub-picosecond stability level}
%
% subtitle is optional
%
%%%\subtitle{Do you have a subtitle?\\ If so, write it here}

\author{\firstname{Konstantin} \lastname{Popov}\inst{1, 2}\fnsep\thanks{\email{popovkon@post.kek.jp} This article will be submitted as a proceedings of the International Workshop on Future Linear Colliders 2025 (LCWS2025)} \and
        \firstname{Hiroshi} \lastname{Kaji}\inst{1, 2} \and
        \firstname{Tetsuya} \lastname{Kobayashi}\inst{1, 2}\and
        \firstname{Aurélien} \lastname{Martens}\inst{3}\and \\
        \firstname{Daniel} \lastname{Charlet}\inst{3}\and
        \firstname{Cedric} \lastname{Esnault}\inst{3}\and
        \firstname{Antoine} \lastname{Back}\inst{3}\and
        \firstname{Paul-Eric} \lastname{Pottie}\inst{4}\and
        \firstname{Fabian} \lastname{Zomer}\inst{3}\and \\
        \firstname{Alexander} \lastname{Aryshev}\inst{1, 2} }

\institute{High Energy Accelerator Research Organization (KEK), 305-0801 Tsukuba, Japan \and The Graduate University for Advanced Studies (SOKENDAI), 240-0193 Kanagawa, Japan
\and IJCLab, CNRS/IN2P3, Université Paris-Saclay, 91405 Orsay, France \and Laboratoire Temps Espace (LTE), Observatoire de Paris, Université PSL, CNRS, Sorbonne Université, 75014  Paris, France}

\abstract{%
A precise synchronization between laser pulse and electron beam arrival time is essential for achieving sub-picosecond stability in modern accelerator facilities. In this work, a Low-Level RF system architecture combined with White Rabbit based timing system has been tested through a collaboration between KEK (Japan) and CNRS/IN2P3, IJClab (France). The setup combines a frequency standard generator, an IDROGEN carrier board with an embedded White Rabbit node, and SkyWorks synthesizers of different form factors to distribute phase-locked clock signals over telecommunication fiber. Phase noise power spectral density measurements were performed at several RF sub-harmonics to confirm synchronization performance. These results demonstrate the feasibility of implementing the White Rabbit–IDROGEN synchronization scheme for large-scale accelerators, including applications to laser-based diagnostics.
}
\maketitle
\section{Introduction}
\label{intro}
The synchronicity between the laser pulse and the beam arrival time is a key requirement for laser-based diagnostics such as polarimetry \cite{Polarimetry_1, Polarimetry_2}, transverse beam size measurements \cite{ATF_IPBSM}, laser-Compton X-Ray generation experiments \cite{KEK_LUCX_proceed, KEK_LUCX_LCS, LLRF_10fs_RMS} and beam position monitor frontend electronics \cite{cBPM_1, cBPM_2}. Therefore, maintaining a stable synchronization between the laser system, the accelerating field, and the overall accelerator timing trigger signals network is of primary importance. In most cases, a reference clock signal—typically a sub-harmonic of the main accelerating frequency—is employed to drive the laser system’s piezo feedback control.
In large-scale accelerator facilities, laser-based diagnostics are often installed hundreds of meters to several kilometers away from the master Low-Level Radio Frequency (LLRF) Signal Generator (SG) \cite{Spring8_LLRF, NSRRC_LLRF_ref}. Conventional RF cable links are unsuitable for such distances due to their sensitivity to electromagnetic interference, temperature variation, and humidity-induced phase drift. Consequently, clock distribution via optical fiber is preferred.
Three main approaches exist for optical fiber–based clock delivery. The first relies on straightforward electro-optical (E/O) and opto-electrical (O/E) conversion without active delay drift compensation \cite{Popov_ATF_LLRF, Popov_ATF_LLRF_and_Timing}. The second introduces feedback-based delay stabilization to mitigate environmental effects \cite{LLRF_10fs_RMS}. The third, increasingly adopted in modern timing systems, employs event-code transmission between a carrier master board  and a local synthesizer board \cite{Spring8_WRS_clock}.
This last approach can be integrated into an event-based timing system architecture \cite{Popov_ATF_timing, Kaji_ATF_timing, Kaji_WRS_SuperKEKB}.
When implemented using the White Rabbit (WR) transmission protocol, it inherently provides precise delay compensation and deterministic phase alignment. In this context, the IDROGEN carrier board  \cite{Popov_IDROGEN}, equipped with an embedded WR node, provides a modular and scalable realization of this concept. As a result, it simplifies LLRF and timing architectures in large accelerator systems and enables low-jitter signal distribution over long distances while reducing complexity and maintenance effort.
\begin{figure}[b]
\centering
\includegraphics[width=12.99cm,clip]{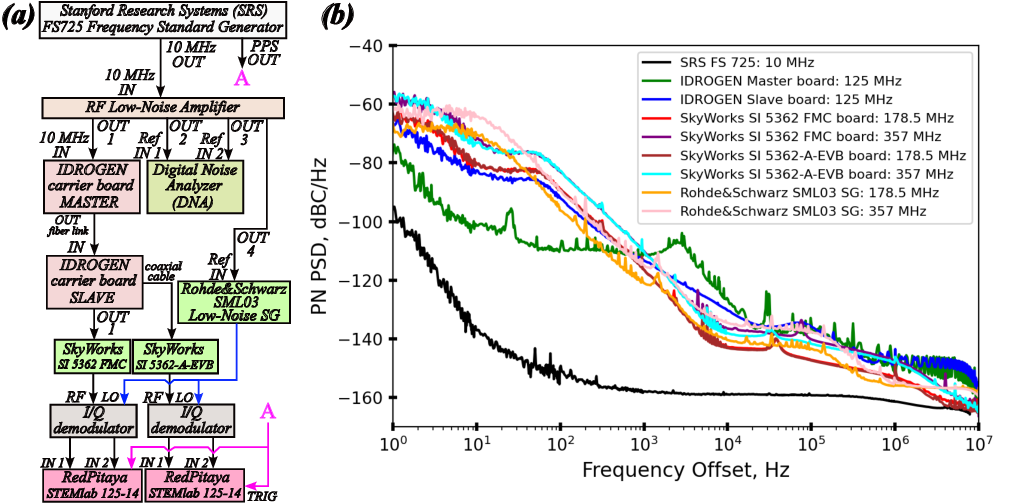}
\caption{The synchronization performance measurement setup: (a) is the block diagram, (b) is the signals PN-PSDs}
\label{fig-1}       % Give a unique label
\end{figure}
\section{IDROGEN carrier board and its synthesizer}
The IDROGEN carrier board, developed at IJCLab, is a µTCA.4 platform based on an Altera FPGA that integrates a fully functional WR timing node \cite{IDROGEN_hardware}. It incorporates a dual-loop PLL clock-jitter cleaner and an in-FPGA Digital Dual-Mixer Time-Difference (DDMtD) phase comparator, providing deterministic and low-jitter clock distribution. The board supports FMC synthesizer mezzanines, such as those built around the SkyWorks SI 5362, enabling precise generation of RF sub-harmonics required for particle accelerator facilities LLRF systems. To confirm the applicability of the carrier board, tests were conducted at IJCLab, KEK ATF, and SuperKEKB. Building on the initial synchronization study reported in  \cite{Popov_IDROGEN}, a detailed jitter and long-term stability analysis was performed at RF frequencies relevant for laser-based diagnostics.
\subsection{Low-level test at CNRS/IN2P3, IJClab}
The IDROGEN master configuration was characterized using the I/Q-demodulation technique \cite{Popov_ATF_timing,  Popov_IDROGEN, ATF_Digitzer_IPAC2025, ATF_Digitzer_PASJ2024}, and the corresponding phase-noise power spectral densities \mbox{(PN-PSDs)} were measured with a Digital Noise Analyzer (DNA).
The master configuration consisted of the Stanford Research Systems (SRS) FS725 frequency standard generator combined with an RF low-noise amplifier (LNA), the IDROGEN master and slave carrier boards, a SkyWorks SI5362-A-EVB synthesizer and  FMC format SI 5362 synthesizer. The phase-detection and acquisition modules are I/Q demodulators with RedPitaya STEMlab 125-14 FPGA board based digitizers \cite{ATF_Digitzer_IPAC2025, ATF_Digitzer_PASJ2024} (see Fig. 1(a)). Both synthesizer boards share the same SI 5362 PLL/VCO core and similar loop parameters, so the PN-PSD results obtained on the SI 5362-A-EVB board are representative for the FMC-based module.
The frequency standard served as the grandmaster, distributing 10 MHz reference to synchronize the IDROGEN master-slave boards chain, the Rohde\&Schwarz SML03 SG, and the DNA.
Once the master was phase-locked, the WR protocol was transmitted over a 25 m optical link to the slave, ensuring deterministic synchronization of both boards.
The synthesizer output phases were then aligned to the SG output ones.
To evaluate the oscillator and PLL stability of the carrier boards, PN-PSD measurements were performed for KEK ATF LLRF system architecture sub-harmonics, specifically the 8th (357 MHz) and 16th (178.5 MHz) sub-harmonics of the 2.856 GHz accelerating frequency \cite{Popov_ATF_timing, Popov_ATF_LLRF}. The offset-frequency range spanned 1 Hz – 10 MHz (see Fig. 1(b)).
\begin{figure}[h]
\centering
\includegraphics[width=13.0cm,clip]{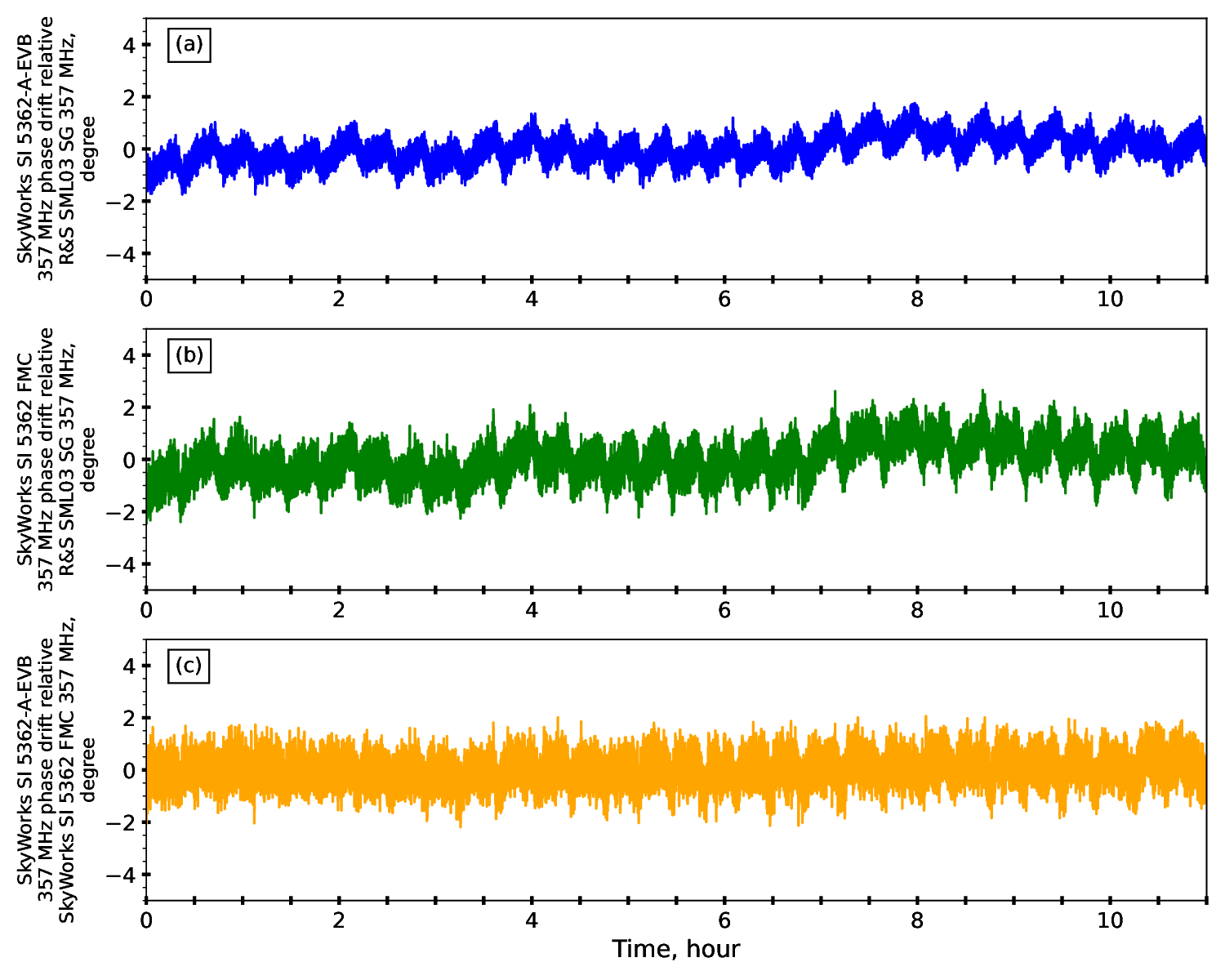}
\caption{The phase stability measurement results: (a) is the SkyWorks SI 5362-A-EVB synthesizer relative Rohde\&Schwarz SML03 SG phase drift vs time, (b) is the SkyWorks SI 5362 FMC synthesizer relative Rohde\&Schwarz SML03 SG phase drift vs time, (c) is the SkyWorks SI 5362-A-EVB synthesizer relative the SkyWorks SI 5362 FMC synthesizer phase drift vs time}
\label{fig-1}       % Give a unique label
\end{figure}

Long-term phase stability was also evaluated for 357 MHz signal over a 12-hour period to assess suitability as a local-oscillator (LO) source for accelerator facilities LLRF systems.
The SkyWorks SI 5362-A-EVB and FMC-based synthesizer demonstrated 3$^{\circ}$ (pp) and 4$^{\circ}$ (pp) relative phase jitter with respect to the SML03 SG, respectively. The I/Q demodulation technique allows phase measurements with 0.1$^{\circ}$ precision \cite{Popov_PhD_thesis}.
Figure 2(a)–(b) show periodic phase fluctuations with approximately 30 min cycles and 1$^{\circ}$ amplitude, superimposed on the slower drift.
For the SI 5362-A-EVB vs FMC comparison (see Fig. 2(c)), the phase-jitter component clearly dominates over the slow fluctuation, confirming good short-term coherence between synthesizers.

The PN-PSD spectra (see Fig. 1(b)) confirm ps-level phase stability of the IDROGEN–SkyWorks synthesizers chain under laboratory condition.
For the 178.5 MHz signal, the phase-noise levels reached –80 dBc/Hz, –110 dBc/Hz, and –140 dBc/Hz at 100 Hz, 1 kHz, and 10 kHz offset, respectively.
At 357 MHz, the PN-PSD is approximately 6 dB higher across the spectrum, consistent with the 6 dB theoretical scaling of phase noise with doubled carrier frequency.
\begin{table}[h]
  \centering
  \footnotesize
  \caption{RMS timing jitter of the signals}
  \label{tab:jitter_only}
  \begin{tabular}{ccccc}
    \hline
    Signal & Frequency& Integration range & RMS jitter$^{a}$ & RMS jitter$^{b}$ \\
    \hline
    SRS FS725 & 10 MHz & 1 Hz -- 5 MHz  & 0.44 ps & 0.44 ps \\
     &  & 10 Hz -- 5 MHz & 0.39 ps & 0.39 ps  \\
    \hline
    IDROGEN Master & 125 MHz   & 1 Hz -- 10 MHz  & 0.54 ps & 0.54 ps  \\
     carrier board & & 10 Hz -- 10 MHz & 0.49 ps & 0.49 ps \\
    \hline
   IDROGEN Slave & 125 MHz & 1 Hz -- 10 MHz  & 1.29 ps & 1.29 ps  \\
    carrier board & & 10 Hz -- 10 MHz & 0.91 ps & 0.91 ps  \\
    \hline
      SkyWorks & 178.5 MHz & 1 Hz -- 10 MHz  & 1.44 ps & 1.44 ps  \\
      SI 5362-A-EVB & & 10 Hz -- 10 MHz & 0.85 ps & 0.85 ps  \\
    \hline
   SkyWorks & 178.5 MHz & 1 Hz -- 10 MHz  & 1.54 ps & 1.41 ps  \\
     SI 5362 FMC & & 10 Hz -- 10 MHz & 0.86 ps & 0.86 ps  \\
   \hline
      SkyWorks & 357 MHz & 1 Hz -- 10 MHz  & 1.49 ps & 1.49 ps \\
      SI 5362-A-EVB & & 10 Hz -- 10 MHz & 0.87 ps & 0.87 ps  \\
    \hline
   SkyWorks & 357 MHz & 1 Hz -- 10 MHz  & 1.54 ps & 1.54 ps \\
     SI 5362 FMC & & 10 Hz -- 10 MHz & 0.87 ps & 0.87 ps  \\
    \hline
    Rohde\&Schwarz & 178.5 MHz & 1 Hz -- 10 MHz  & 1.83 ps & 1.83 ps  \\
    SML03 SG  & & 10 Hz -- 10 MHz & 1.18 ps  & 1.18 ps \\
    \hline
    Rohde\&Schwarz & 357 MHz & 1 Hz -- 10 MHz  & 1.82 ps & 1.82 ps  \\
    SML03 SG  & & 10 Hz -- 10 MHz & 1.21 ps & 1.21 ps  \\
    \hline
  \end{tabular}
  
   \begin{flushleft}  
 \footnotesize   
  $^{a}$ the PN-PSD spikes contribution is taken into account \\
  $^{b}$ the PN-PSD spikes contribution is not taken into account \\
 \end{flushleft}  
  
\end{table}
The IDROGEN slave-board output at 125 MHz reproduced the master-board noise up to 1 kHz offset, after which its phase noise followed the internal PLL behavior of the master.
Overall, the measured spectral hierarchy—from the SRS FS725 reference to the IDROGEN carrier and finally to the FMC and SkyWorks synthesizer outputs—shows that the system’s performance is primarily limited by the quality of the 10 MHz reference signal and by the phase-locked-loop (PLL) parameters of the IDROGEN master board.
For reference, the SRS FS725 10 MHz output itself exhibits –100 dBc/Hz at 1 Hz, –140 dBc/Hz at 10 Hz, –150 dBc/Hz at 100 Hz, and –160 dBc/Hz at 1 kHz, establishing the baseline used for all PN-PSD normalization.
Integrating the measured spectra yields timing jitter below \mbox{1 ps rms} (10 Hz–10 MHz), confirming that the WR-disciplined IDROGEN system can deliver the sub-picosecond stability required for laser-to-beam synchronization at modern accelerator facilities. Spurious narrowband spikes in the PN PSD were removed using a median-filter-based despiking algorithm. 
These laboratory measurements established the baseline performance of the IDROGEN carrier board with the SkyWorks SI 5362 synthesizer chain. To verify its applicability in a realistic accelerator environment, the system next tests were conducted at the KEK ATF LLRF and SuperKEKB setups.
\begin{table}[b]
  \centering
  \footnotesize
  \caption{RMS timing jitter of the signals}
  \label{tab:jitter_only}
  \begin{tabular}{ccccc}
    \hline
    Signal & Frequency& Integration range & RMS jitter$^{a}$ & RMS jitter$^{b}$ \\
    \hline
    SRS FS725 & 10 MHz   & 1 Hz -- 5 MHz  & 0.44 ps& 0.44 ps \\
      & & 10 Hz -- 5 MHz & 0.39 ps& 0.39 ps \\
    \hline
   IDROGEN Master & 100 MHz & 1 Hz -- 10 MHz  & \textbf{2.31 ps} & \textbf{0.77 ps}\\
    carrier board & & 10 Hz -- 10 MHz & \textbf{2.31 ps} & \textbf{0.76 ps}\\
    \hline
   IDROGEN Slave & 250 MHz & 1 Hz -- 10 MHz  & 1.13 ps& 1.12 ps \\
    carrier board & & 10 Hz -- 10 MHz & 0.91 ps & 0.89 ps\\
    \hline
      SkyWorks & 178.5 MHz & 1 Hz -- 10 MHz  & 1.50 ps& 1.49 ps \\
      SI 5362 FMC & & 10 Hz -- 10 MHz & 0.89 ps& 0.88 ps \\
    \hline
     SkyWorks & 357 MHz & 1 Hz -- 10 MHz  & 1.43 ps& 1.42 ps \\
      SI 5362 FMC & & 10 Hz -- 10 MHz & 0.88 ps& 0.87 ps \\
     \hline
  \end{tabular}
\begin{flushleft}  
 \footnotesize   
  $^{a}$ the PN-PSD spikes contribution is taken into account \\
  $^{b}$ the PN-PSD spikes contribution is not taken into account \\
 \end{flushleft}  
\end{table}
\subsection{Low-level test at KEK ATF}
The second test was conducted at the KEK ATF LLRF system \cite{Popov_ATF_LLRF} to confirm the \mbox{IDROGEN} carrier board applicability for the International Linear Collider beam instrumentation R\&D program.  The SRS FS725 reference outputs were injected into the Agilent E8663B master SG and the IDROGEN master board (see Fig. 3(a)). 
\begin{figure}[h]
\centering
\includegraphics[width=13cm,clip]{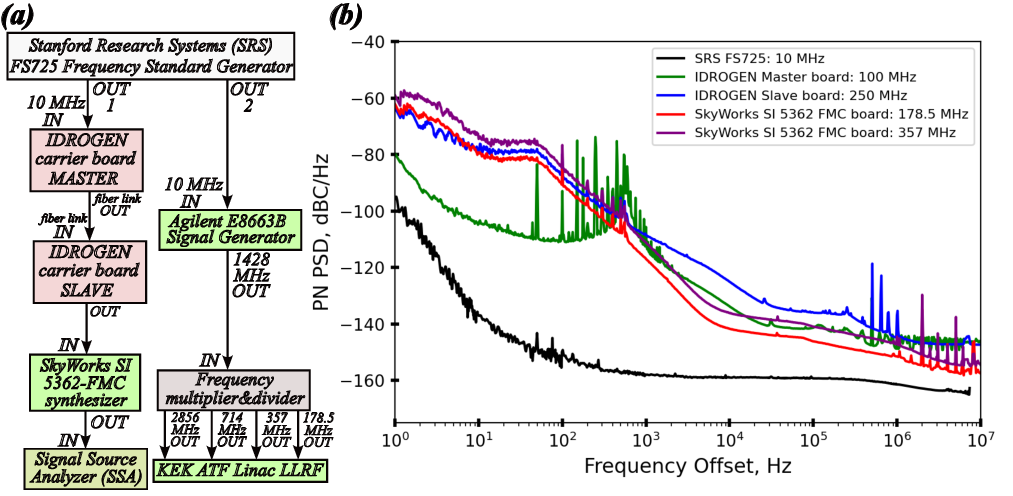}
\caption{The synchronization performance measurement setup: (a) is the block diagram, (b) is the signals PN-PSDs}
\label{fig-1}       % Give a unique label
\end{figure}
Then, the IDROGEN master and slave boards output signals were aligned with respect to the input 10 MHz. As a result, the Agilent E8663B master SG and synthesizer connected to the IDROGEN slave board were synchronized with each other. The SkyWorks SI 5362-A-EVB synthesizer board output signal frequency was 178.5 MHz, which is the KEK ATF RF-Gun laser oscillator pulses repetition rate.
The PN-PSDs associated with the signals are shown on the Figure 3(b). It has to be mentioned that the frequency modulation of the SG was turned off \cite{Popov_ATF_LLRF}.  The Agilent E8663B SG output is 1428 MHz. On the next stage, the signal was injected into the multichannel frequency divider\&multiplier module. There are 4 output signals from the module. These are 178.5 MHz, 357 MHz, 714 MHz and 2856 MHz \cite{Popov_ATF_LLRF, Popov_ATF_timing}. The PN-PSD data was analyzed for 2 cases. The first one takes the PN-PSD spikes (see Fig. 3(b)) contribution into account, while the second case does not include it. As can be seen on the Figure 3(b), the IDROGEN master board PN-PSD has several spikes located between 50 Hz and 3 kHz. These increase IDROGEN master carrier board  RMS jitter value from \textbf{0.77 ps} to \textbf{2.31 ps}. However, \mbox{the SkyWorks SI 5362 FMC} synthesizer PN-PSD spikes do not affect the RMS jitter (see Table 2).

\subsection{Low-level test at SuperKEKB}
The synchronization test of the IDROGEN carrier board within an accelerator LLRF system was conducted at SuperKEKB \cite{SuperKEKB_LLRF_1}.
The first IDROGEN board was configured as the master, while the second operated as the slave (see Fig. 4(a)).
In the SuperKEKB LLRF system, an Agilent E8663D signal generator produced a 510 MHz sine-wave that was frequency-divided by 51 to generate a 10 MHz reference for the entire system, including the IDROGEN boards.
As a result, the 510 MHz generator acted as the grandmaster reference. The second Agilent E8663D generated the 508.9 MHz signal used by the HER and LER LLRF branches. 
\begin{figure}[htb]
\centering
\includegraphics[width=13cm,clip]{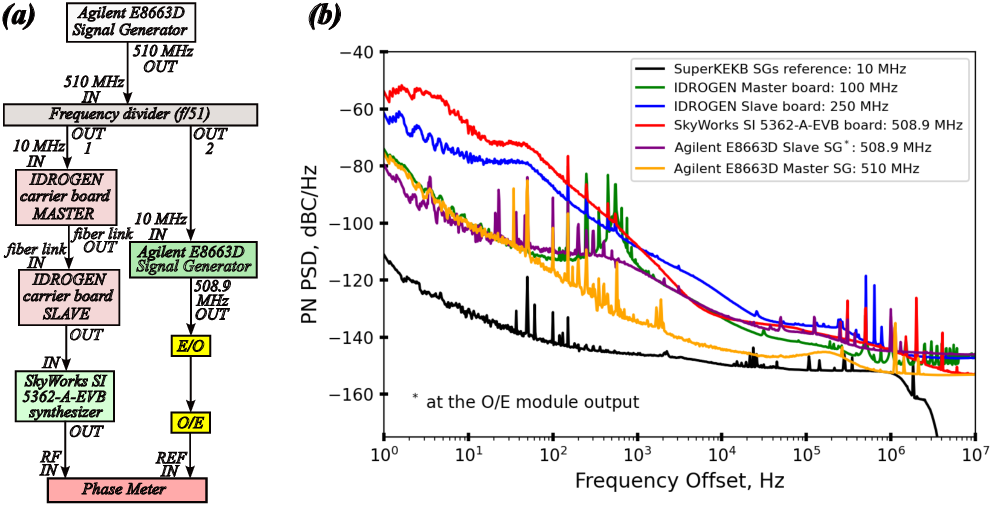}
\caption{The synchronization performance measurement setup: (a) is the block diagram, (b) is the signals PN-PSDs}
\label{fig-1}       % Give a unique label
\end{figure}
At the next stage, a SkyWorks SI 5362 FMC synthesizer generated a 508.9 MHz clock synchronized with the HER/LER SG output \cite{SuperKEKB_LLRF_1, SuperKEKB_LLRF_2}. Both signals were fed into the phase meter module based on the I/Q demodulation to record long-term synchronization-stability data. In this configuration, the IDROGEN master board disciplined the slave board through the WR protocol, while the SkyWorks synthesizers derived their reference from the slave carrier board output. The long-term synchronization stability was measured for 12 hours (see Fig. 5).  The SkyWorks SI 5362 FMC-based synthesizer demonstrates 4$^{\circ}$ (pp) and 0.61$^{\circ}$ (RMS) phase stability with respect to the HER/LER SG LLRF branch over 12 hours.  In terms of synchronization stability, these values are equal to the 21.8 ps (pp) and 3.3 ps (RMS).  The long-term phase stability was therefore evaluated on this FMC implementation, while the PN-PSD corresponds to the SI 5362-A-EVB evaluation board.
The PN PSDs were measured by an Agilent E5052B Signal Source Analyzer (SSA). The SSA measurement settings were 16 averages and 2 correlations. 
The SuperKEKB LLRF 10 MHz reference exhibited phase-noise levels of approximately –110 dBc/Hz, –130 dBc/Hz, and –140 dBc/Hz at 1 Hz, 10 Hz, and 100 Hz offset, respectively (see Fig. 4(b)).
Beyond \mbox{100 Hz}, the phase-noise level gradually decreased from –140 dBc/Hz to –150 dBc/Hz at \mbox{1 MHz}.
\begin{figure}[htb]
\centering
\includegraphics[width=13cm,clip]{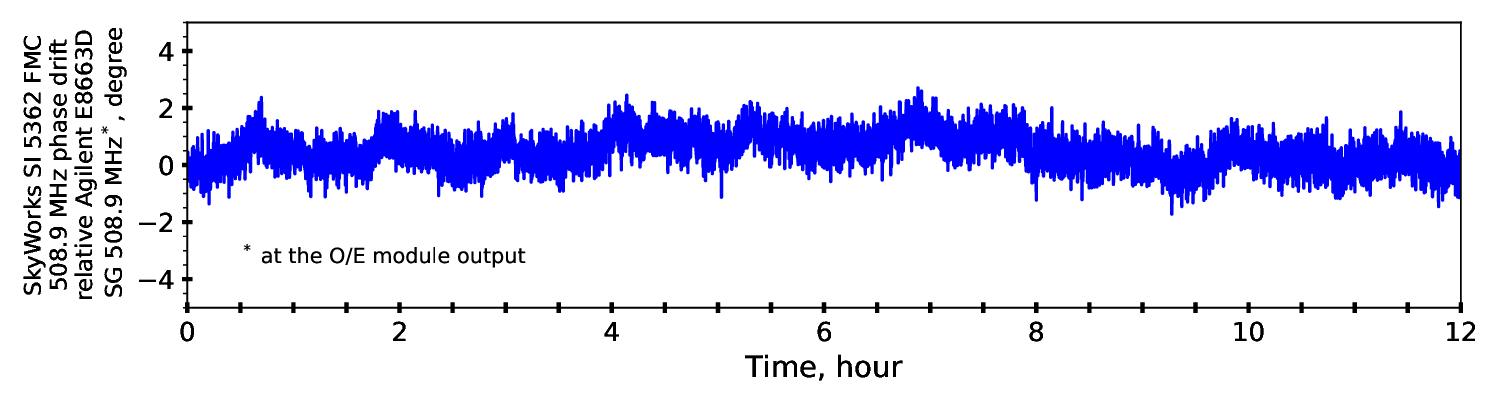}
\caption{The phase stability measurement results}
\label{fig-1}       % Give a unique label
\end{figure}
The IDROGEN boards operating at 100 MHz and 250 MHz displayed phase-noise behavior consistent with the measurements reported in Section 2.1 and 2.2.
Meanwhile, the Agilent E8663D signal generators delivering 510 MHz and 508.9 MHz outputs demonstrated outstanding phase-noise performance, reaching –80 dBc/Hz, –100 dBc/Hz, and –105 dBc/Hz at 1 Hz, 10 Hz, and 100 Hz offset, respectively. 
Moreover, enabling frequency modulation on the 510 MHz grandmaster SG did not degrade the phase-noise level or the HER/LER SG output synchronization quality with the SkyWorks SI 5362-A-EVB synthesizer \cite{SuperKEKB_LLRF_2}.
\begin{table}[h]
  \centering
  \footnotesize
  \caption{RMS timing jitter of the signals}
  \label{tab:jitter_only}
  \begin{tabular}{ccccc}
    \hline
    Signal & Frequency& Integration range & RMS jitter$^{a}$ & RMS jitter$^{b}$ \\
    \hline
    SuperKEKB SGs & 10 MHz & 1 Hz -- 5 MHz  & 0.75 ps & 0.73 ps \\
     reference &  & 10 Hz -- 5 MHz & 0.75 ps & 0.73 ps \\
    \hline
    IDROGEN Master & 100 MHz   & 1 Hz -- 10 MHz  & \textbf{1.11 ps} & \textbf{0.65 ps} \\
     carrier board & & 10 Hz -- 10 MHz & \textbf{1.06 ps} & \textbf{0.56 ps} \\
    \hline
   IDROGEN Slave & 250 MHz & 1 Hz -- 10 MHz  & 1.26 ps & 1.25 ps \\
    carrier board & & 10 Hz -- 10 MHz & 0.92 ps & 0.90 ps \\
    \hline
      SkyWorks & 508.9 MHz & 1 Hz -- 10 MHz  & 2.07 ps & 2.07 ps \\
      SI 5362-A-EVB & & 10 Hz -- 10 MHz & 0.94 ps & 0.94 ps \\
    \hline
  Agilent E8663D & 508.9 MHz & 1 Hz -- 10 MHz  & 0.11 ps & 0.09 ps\\
     Slave SG & & 10 Hz -- 10 MHz & 0.10 ps & 0.09 ps \\
    \hline
  Agilent E8663D & 510 MHz & 1 Hz -- 10 MHz  & 0.08 ps & 0.07 ps \\
    Master SG  & & 10 Hz -- 10 MHz & 0.05 ps & 0.04 ps \\
    \hline
  \end{tabular}
  
 \begin{flushleft}  
 \footnotesize   
  $^{a}$ the PN-PSD spikes contribution is taken into account \\
  $^{b}$ the PN-PSD spikes contribution is not taken into account \\
 \end{flushleft}  

\end{table}
In general, the PN-PSD spikes do not significantly contribute to the signals RMS noise, except the IDROGEN master carrier board one. Its RMS noise with spikes contribution is equal to \textbf{1.11 ps} and \textbf{1.06 ps} for 1 Hz to 10 MHz and 10 Hz to 10 MHz integration ranges, respectively.  However, if the spikes contribution is not taken into account, the RMS noise is equal to \textbf{0.65 ps} and \textbf{0.56 ps} for 1 Hz to 10 MHz and 10 Hz to 10 MHz integration ranges, correspondingly.

\section{Conclusion}
In this work, the White-Rabbit–disciplined IDROGEN carrier board combined with \mbox{SI 5362-based} synthesizers was experimentally evaluated  as a candidate synchronization platform for large-scale accelerator LLRF systems and laser-based diagnostics. Tests at IJCLab, KEK ATF and SuperKEKB show that the WR–IDROGEN chain can distribute RF reference signals 178.5, 357 and 508.9 MHz with sub-picosecond to few-picosecond rms jitter for the 10 Hz–10 MHz integration range and long-term relative phase drift of about  \mbox{3$^{\circ}$ – 4$^{\circ}$ pp} at 357 MHz (23.3 ps – 31.1 ps pp)  and 4$^{\circ}$ pp (21.8 ps pp) at 508.9 MHz over 12 h.
These results demonstrate that the WR-based IDROGEN scheme is a viable solution for delivering low-jitter local-oscillator signals to remote laser systems, beam diagnostics and LLRF frontend electronics, including frequency down- and up-conversion stages, in modern accelerator facilities. Future work will address operation over km-scale links, mitigation of narrowband spurs in the phase-noise spectrum, and demonstration of closed-loop laser-to-beam arrival-time stabilization for applications such as Compton polarimetry and nanometre-scale beam-size monitors.
\section{Acknowledgement}
The research leading to these results received funding from the European Union’s Horizon Europe MSCA programme under grant agreement No. 101086276.
Dr. Konstantin Popov was supported by the KEK “Support Program for Long-term Overseas Research.” The authors are also grateful for support from TYL-FJPPN.
This work was additionally supported by the MEXT programm “Development of key element technologies to improve the performance of future accelerators,” Japan Grant Number JPMXP1423812204. Part of the instrumentation was funded by the “Investissements d’Avenir” programme launched by the French Government and implemented by the Agence Nationale de la Recherche (ANR) under reference ANR-21-ESRE-0029 \mbox{(ESR/Equipex+ T-REFIMEVE)}.
%
% BibTeX or Biber users please use (the style is already called in the class, ensure that the "woc.bst" style is in your local directory)
% \bibliography{name or your bibliography database}
%
% Non-BibTeX users please use
%

\end{document}